# Multi-value Probabilistic Computing with current-controlled Skyrmion Diffusion

*Thomas B. Winkler[1*], Yuean Zhou[2], Grischa Beneke[2], Fabian Kammerbauer[2], Sachin Krishnia[2], Mario Carpentieri[3], Davi R. Rodrigues[3§], Mathias Kläui[2,4#], Johan H. Mentink[1+]*

1. Radboud University, Institute for Molecules and Materials, Nijmegen 6525, the Netherlands
2. Institut für Physik, Johannes Gutenberg-Universität Mainz, Mainz 55099, Germany
3. Department of Electrical and Information Engineering, Politecnico di Bari, Bari, Italy
4. Center for Quantum Spintronics, Department of Physics, Norwegian University of Science and Technology, Trondheim 7491, Norway

\* thomas.winkler@ru.nl, § davi.rodrigues@poliba.it + johan.mentink@ru.nl # klaeui@uni-mainz.de

## Abstract

Magnetic systems are highly promising for implementing probabilistic computing paradigms because of the fitting energy scales and conspicuous non-linearities. While conventional binary probabilistic computing has been realized, implementing more advantageous multi-value probabilistic computing (MPC) remains a challenge. Here, we report the realization of MPC by leveraging the thermally activated diffusion of magnetic skyrmions through an effectively non-flat energy landscape defined by a discrete number of pinning sites. The time-averaged spatial distribution of the diffusing skyrmions directly realizes a discrete probability distribution, which is tunable by current-generated spin-orbit torques, and can be quantified by non-perturbative electrical measurements. Even a very straightforward implementation with global tuning, already allows us to demonstrate the softmax computation – a core function in artificial intelligence. As a key advance, we demonstrate invertible logic without the need to create a network of probabilistic devices, offering major scalability advantages. Our proof of concept can be generalized to multiple skyrmions and can accommodate multiple locally tunable inputs and outputs using magnetic tunnel junctions, potentially enabling the representation of highly complex distribution functions.

# Introduction

Over the past decade, computational demands have surged (De Vries, 2023), largely driven by the rapid advancement of artificial intelligence (AI). Both for economic and ecological concerns, technological advancements in both software and hardware are being pursued to support the increasing demands of AI algorithms. One route to achieving higher hardware efficiency is to increase task parallelism using hardware accelerators such as FPGAs, MRAMs, or ASICs (Capra et al., 2020) or by enhancing information density at the device level, an approach that can be realized through analog computing (Finocchio et al., 2024). A key example of the latter is the implementation of probabilistic computing paradigms, which harness the inherent noise in hardware to perform stochastic operations - eliminating the overhead of having to generate random numbers digitally (Borders et al., 2019). One of the prevailing paradigms in probabilistic computing is the probabilistic bit (p-bit), which can be implemented using stochastic magnetic tunnel junctions (sMTJs) (Zink et al., 2022). These devices have the potential to replace thousands of conventional transistors with a single element and can be integrated in a massively parallel fashion using FPGA technology (Aadit et al., 2022; Yang et al., 2025). However, the performance of conventional sMTJ-based p-bit- systems may be constrained by device imperfections and variability between individual components (Aadit et al., 2022; Raimondo et al., 2025). Furthermore, the binary nature of p-bits requires creating complex networks of p-bits to perform complex operations in, for example, AI and optimization problems, posing considerable challenges in terms of scalability and energy efficiency (Soh et al., 2025).

An appealing route to bypass such challenges is to further enhance the information density using multi-value probabilistic computing (MPC) elements. Unlike binary bits, which represent only two states (0 and 1), MPC elements can encode multiple discrete states, offering higher data density and reduced hardware redundancy (Andreev et al., 2022). MPC plays a crucial role in handling uncertainty, a cornerstone challenge in AI and related advanced computational models (Nikhar et al., 2024). Traditional binary logic, which constrains representation to strict true or false values, often falls short when dealing with imprecision, incomplete data, or ambiguity (Celikyilmaz, Türksen, 2009). In contrast, MPC enables information encoding over a range of values, enabling a natural representation of uncertainty (Gawlikowski et al., 2023; Zadeh, 2023) and is relevant in various fields of research, such as in fuzzy and invertible logic, stochastic neural networks, and Bayesian computation (Gawlikowski et al., 2023; Zadeh, 1988, 1999; Beaumont, 2019; Parr et al., 2018; Rohmer, 2020; Rue et al., 2017; Onizawa et al., 2021; Smithson et al., 2019; Camsari et al., 2017; Jaber et al., 2020, Lukasiewicz 1999).

Despite the relevance and potential advantages, the on-device realization of MPC presents significant challenges (Yoo et al., 2021; Andreev et al., 2022). Existing efforts in spintronics have largely focused on integrating conventional p-bit architectures (Borders et al., 2019; Camsari et al., 2017; Jia et al., 2020; Aadit et al., 2021; Xia & Yang, 2019), which faces similar scalability challenges as for creating networks of p-bits. Generalizations of p-bits have been proposed, such as probabilistic digits (p-dits) and Gaussian bits (g-bits) (Debashis et al., 2021; Nihal Sanjay Singh et al., 2024; Duffee et al., 2025), which are typically constructed by cascading multiple p-bits (W. Whitehead, Z. Nelson, Camsari, 2023). However, their application so far has largely focused on Potts models (Wu, 1982), which still require creating networks of probabilistic elements and hence only partially leverage the full expressive power of MPCs (W. Whitehead, Z. Nelson, Camsari 2023).

Here, we introduce a conceptually novel approach to MPC that does not rely on creating networks of probabilistic elements and therefore offers significant scalability advantages. Our approach leverages the stochastic dynamics of magnetic skyrmions in thin films - nanoscale spin textures that undergo Brownian motion and thereby explore the effective energy landscape of the system. The energy landscape includes areas where the skyrmion has a lower energy. Such local minima in the energy landscapes are sometimes termed pinning sites, which can emerge from imperfections during the fabrication process (Gruber et al., 2022). However, these sites can also be intentionally engineered during fabrication, enabling precise tuning of the local energy profile (Kern et al., 2022; He et al., 2023). While previous studies regarded pinning as a detrimental effect to be mitigated through thermal activation and applied currents (Raab et al., 2022; Gruber et al., 2023; Beneke et al., 2024), our approach embraces these pinning sites as essential functional states for computation. As the time spent by magnetic skyrmions in the pinning sites follows the Boltzmann distribution (Brems et al., 2025), we use the collective ensemble of pinning sites directly to represent the number of discrete states. The probabilistic distribution across these states is obtained by time-multiplexing the trajectories of thermally activated skyrmions (Brems et al., 2021; Raab et al., 2022; Beneke et al., 2024). Moreover, we harness the tunability of the local energy of the pinning site by current-induced spin-orbit torques. We focus specifically on current-driven tuning of the energy landscape, demonstrating that a single applied current can effectively engineer the system to realize multiple operational scenarios.

To demonstrate the relevance of our MPC concept for probabilistic computing, we consider two judiciously chosen key problems. Firstly, we utilize the similarity between the Boltzmann distribution and the softmax function to demonstrate the physical realization of a diffusion-based softmax distribution, which is a key computational task in modern generative models and various types of classifiers, such as transformers (Vaswani et al., 2017). We show simulation evidence for a successful training with diffusion-based softmax. Secondly, we realize an inverted OR-gate with the skyrmion system experimentally. Invertible logic is a method to tackle NP-complete combinatorial optimization problems, including Max-SAT (Grimaldi et al., 2022). We reach a competitive performance of the invertible logic gate quality when compared to realizations implemented with multiple p-bits. Moreover, due to an exact mapping of only desired truth table entries to pinning sites, we can drastically reduce the sampling space, which significantly reduces degeneracy issues and maximizes the sampling accuracy as compared to implementations based on Ising machines, which sample the entire state space of needed variables (Aadit, 2022). In the skyrmion diffusion case, each entry in the truth table is mapped to a pinning site, which is done once during the calibration phase. This not only demonstrates that a single skyrmion diffusion system can represent a multi-value distribution, but also shows that invertible logic is possible without the need for dense interconnects, cascading architectures, and stringent control over device uniformity, thereby significantly enhancing scalability and reducing energy demands.

## **Current controlled skyrmion diffusion in a non-flat effective energy landscape**

In order to implement MPC we first characterize and model our sample. We analyze skyrmion dynamics in a confined geometry, using thin-film deposited materials stabilizing skyrmions by interfacial DMI and dipolar fields. Our thin-film stack allows for smooth amorphous growth of the ferromagnetic material and leads to a comparatively flat effective energy landscape, allowing for thermally activated skyrmion dynamics at roughly room temperature (details, see Methods). However, due to unavoidable local material inhomogeneities such as grain boundaries in some layers, leading to local thickness variations and other growth inhomogeneities and sample edges, the skyrmion dynamics are governed by this given effective energy landscape. Regions of reduced anisotropy or exchange might pin the skyrmion domain walls at these regions, leading to a pinning-dominated regime of diffusion in our systems (Gruber et al., 2023). We call these areas pinning sites, and the number of pinning sites $K$ is naturally given by the (randomly distributed) material imperfections. Based on this, **Figure 1** illustrates the skyrmion-based MPC concept, visualizing the skyrmion as a diffusing particle in its effective energy landscape. Time- and spatially resolved imaging of the magnetization was performed by optical Kerr-microscopy (Schmidt et al., 1985; Schäfer 2007). The typical skyrmion size in our stacks is in the order of one to a few micrometers.

**Figure 2a** shows an exemplary Kerr microscopy snapshot of the magnetization configuration, featuring a triangular-shaped sample with electrical contacts at the edges. The out-of-plane magnetization is revealed by the gray scale contrast within the triangle; the skyrmion appears as a slightly darker area compared to the surrounding. Next we study the energy landscape by measuring what percentage of the time the skyrmion stay at the inhomogeneities. **Figure 2b** shows the cumulative occurrence for five measurements, each taken from a long-time Kerr microscopy measurement (minimum 2,5 hours at 16 Hz frame rate and 320 Kelvin). The measurements clearly reveal $K=6$ pinning sites, and their centers are indicated with colored discs. **Figure 2c-e** shows similar skyrmion occurrence maps for various fixed applied voltages of our triangular samples (between -1.5 mV and +2 mV, specifying the strength of the spin-orbit torque. The current density at half width of the triangle is estimated to be smaller than 1e8 A/m² even for the highest applied voltage). The voltage is applied between the bottom right and the top contact and pushes the skyrmion towards one of the corners with the contacts, depending on the sign and strength of the applied voltage $U$. This is consistent with the expectation that skyrmions move along the electron flow for spin-orbit torques. In the limit of infinitely long measurements, the normalized occurrence map $\pi_{\exp}(U)$ should converge to the stationary distribution $\pi(U)$ of the system that is given by the energy landscape.

## **Markov State Modeling**

Having established the energy landscape, we next explore the dynamics of the skyrmion motion between different pinning sites. From a device perspective, tracking the skyrmion position is crucial for the device functionality. Due to thermal fluctuations, a natural lower limit of the tracking accuracy could be the size of a local pinning site. To analyze dynamical systems with discrete states defined by pinning sites and influenced by thermal fluctuations, Markov State Modeling (MSM) is an approach (Husic and Pande 2018). So we extract the coarse-grained MSM for the appropriate number of pinning sites $K$ using a clustering algorithm. The core of the MSM-based clustering algorithm is the row-stochastic transition matrix $\underline{W}^{(K \times K)}$, where the element $W_{ij}$ describes the conditional probability of a transition from state $i$ to $j$ ($i \rightarrow j$) within the lag time $\tau$, while being in state $i$. In our case, $\tau$ is chosen as the frame rate of the Kerr microscope $1/\tau = 16$ Hz. The clustering procedure coarse-grains the sample area into $K$ pinning sites by maximizing the diagonal elements of $\underline{W}^{(K \times K)}$ (self-transitions) and is described in detail in the Methods section.

Assuming a Boltzmann distribution $\pi_{\exp}(x) = e^{(-(V(x)))}/S$, with the normalization integral $S(V) = \int \left(e^{(-(V(x)))} dx\right)$ and $V(x)$ the potential, one can attribute an energy from a stationary distribution $\pi$. We consider $\beta = 1/(k_B T) = 1$ for simplicity and omit it in the formulae. In a discrete setup, enumerating pinning sites with $i \in \{1, 2, ..., K\}$, we rewrite $S(V) = \sum_i e^{(-V_i)}$. Importantly, as in the pinning-dominated regime and for our large skyrmions with narrow domain walls the skyrmion Hall angle can be neglected, we model the spin-orbit torque as a conservative force, $F_{\text{SOT}} = -\nabla V_{\text{SOT}}$, neglecting also Joule heating. Thus, also systems with a (fixed) applied voltage will fulfill 'detailed balance' and reach a thermodynamic equilibrium. The model for the pinning site energies is thus $E_i(U) = E_i^0 + c_i \cdot U$, with $i$ indicating the respective pinning site. For each pinning site, a scaling factor $c_i$ is fitted to account for the complex current density path. The scaling factor (in theory) depends on the position of the pinning site relative to the contacts. As the numerical minimization process is highly sensitive to the initial values, we provide initial values based on current path simulations (for details, see Methods).

We use modeled energies for each pinning site as a function of the voltage, as shown in **Figure 3a**. Exploiting the initial values based on current path simulations, we find slopes $c_i$ being positive or negative, depending on the position of the pinning site with respect to the contacts and current flow in the triangle. For example, the center pinning site (blue) features a small slope, while the outer pinning sites (green, yellow, brown) have a much steeper slope with changing applied voltage depending on the effect of the spin-orbit torques. **Figure 3b** shows the modeled stationary distribution and its scaling with current, obtained from the calculation of the Boltzmann distribution. We compare the modeled results to experimental data that is marked as triangular points, showing good agreement. There are some minor deviations that can reflect errors due to the tracking procedure or limitations resulting from the assumption of a linear scaling.

# **Skyrmion-diffusion based softmax calculation**

Having demonstrated the possibility of modeling the system in accordance with the Boltzmann distribution, we apply our system to demonstrate the relevance of our MPC for implementing the first functionality. We have chosen the important softmax calculation, which is widely used in neural networks and classification tasks (Banerjee et al., 2020; Han et al., 2024; He et al., 2018; Maharjan et al., 2020; Zhu et al., 2020).

Considering a discrete, categorical variant of the Boltzmann law (with $k_BT=1$), one directly observes that a discretized skyrmion reservoir realizes $e^{-y}/S(-y) = $ softmax($\mathbf{y}$), with $\mathbf{y}=\mathbf{Wx}$ being the pre-activations of the current layer of a neural network (**W** the weights and **x** the input of the current layer, and $S(y)=\sum e^{(-y_i)}$). This allows for direct mapping from pinning site energies **E**(U) to pre-activation inputs. The effective energy landscape experienced by the skyrmion can be divided into multiple regions corresponding to different negative class pre-activations, with the dwell time in each region representing the probability of selecting a given class. This functionality is particularly useful in deep learning accelerators, where efficient softmax implementations are essential for real-time performance. The number of available classes is directly mapped to the number of pinning points for the chosen classification task. Translating the concept to the probabilistic model obtained from experimental data of the diffusive skyrmion system, **Figure 3b** already shows the variety of possible output distributions in our system with six pinning sites. The effective energies, plotted in **Figure 3a**, are for this functionality interpreted as the negative pre-activations.

To apply this concept to a real dataset, we consider the Monte Carlo simulation of a diffusion-based softmax operation. We demonstrate diffusion-based softmax in a small network for the Iris dataset (Fischer, 1936), which is a standard benchmark often used to evaluate machine learning classifiers (Pinto et al., 2018; Swain, 2012). The Iris dataset contains images of three subvariants of the Iris flower that need to be distinguished.

The simulation contains a small neural network (**Figure 4a**) with a softmax activation function that is both evaluated directly using the analytical formula as well as by simulating the diffusing skyrmion as a (quasi-)particle in the Thiele approach (Thiele 1973; Lin et al., 2013). For this analysis approach, the skyrmion is treated as a rigid point particle. To reduce the computational cost, the system dimensions are rescaled. At each time step, the skyrmion motion was determined using the Metropolis-Hastings algorithm, governed by the local energy at each position. A circular sample with a radius of 0.1 was divided into three regions, each of which was assigned to one of the three flower classes (**Figure 4b**). The local energy in each region was used as pre-activations for the softmax function. The pre-activation values were typically within a range of approximately -1 to 1. The output of the diffusion-based softmax function is based on the dwell time within each region, estimated by running the simulation for 100,000 time steps. It is important to note that there is a trade-off between simulation accuracy and computational time. These outputs were used during both the training and inference stages. **Figure 4c** shows the difference between the analytically predicted and statistically derived values (at a random state of the training). To ensure statistical robustness, the classification task was repeated ten times. We find a competitive performance for the diffusion-based solution in the training as well as in the validation set (**Figure 4d**). The diffusion-based solution performs comparably. The discrepancy between the diffusion-based results and the exact softmax calculations increases slightly in the high-accuracy regime, where the pre-activation inputs exhibit larger separations. This behavior arises because larger gaps between input values require longer simulation times to accurately approximate the softmax probabilities through finite sampling.

Our system, by contrast, demonstrates excellent approximation performance when the differences between pre-activation values are small. This behavior is particularly advantageous for applications such as Attention mechanisms (Vaswani et al., 2017), which involve computing large distributions of pre-activations weighted by queries, keys, and values. Notably, the linear operations required in these computations align well with the demonstrated linear energy scaling shown in **Figure 3a**. The ability of the device to efficiently approximate probability distributions over the energy landscape is essential for preserving both computational accuracy and parallelism in such architectures (Maharjan et al., 2020; Vaswani et al., 2017).

## **Invertible Logic Gate with skyrmion diffusion based MPC**

Having demonstrated the performance of the skyrmion-based MPC as an approximate softmax calculator, as the second key functionality, we evaluate the use of diffusive skyrmions in the pinned regime for invertible logic. Invertible logic is one of the most important modern, probabilistic computing paradigm that holds promise in the field of cryptography and for solving for instance complex optimization problems (Camsari, 2017). Conceptually, invertible logic aims to find the corresponding set of input states given an output state. Within MPC it proceeds by clamping one output and then sampling the energy landscape to find the corresponding input states of a given (Boolean) logic gate. Notably, the system simultaneously provides all output values after the sampling time, enabling fast and efficient parallel computation and significantly increasing processing efficiency. Current implementations typically use coupled stochastic magnetic tunnel junctions (Camsari et al., 2017). However, despite the fact that high energetic states tend to be suppressed, the auxiliary coupling matrix still leads to a limited sampling accuracy in experimental realizations (Aadit et al., 2021). In the presented skyrmion system, each pinning site is mapped to a particular entry in the truth table of the target logic gate, avoiding that states not present in the truth table are sampled (equal to 100 % sampling accuracy). This direct mapping is done once during calibration and eliminates invalid logic states at the hardware level, providing a more reliable and deterministic approach than relying solely on suppressing them via high-energy configurations. Logical inputs and outputs are enforced by shaping the energy landscape through the application of targeted voltages. In an ideal scenario where the energies of the pinning sites can be tuned independently, the architecture exhibits linear scaling with respect to logic complexity. This linearity, combined with the elimination of invalid states and integration into a single magnetic structure with minimal wiring, enables fast, parallel, energy-efficient computation, substantially enhancing overall processing performance.

**Figure 5** illustrates the performance of our MPC prototype when used as an invertible OR gate. We perform experiments on a second sample with a clustering into four pinning sites to match the number of entries in a truth table of the OR gate. **Figure 5a** visualizes the stationary distributions (modeled and measured), while **Figure 5b** shows the energy landscape, the approximate pinning site center, and their mapping to the truth table entries. The truth table of the OR-gate contains four different true output states. To find the voltage that best matches $\pi(U)=\pi_{\text{1-clamped}}$, $\pi_{\text{0-clamped}}$ or $\pi_{\text{unclamped}}$, we firstly perform measurements to model the system (triangle in **Figure 5a**). We then scan the voltage and calculate the Kullback-Leibler divergence ($\text{KLD}(U)=\sum_i \pi_{\text{GOAL},i}\log(\pi_{\text{GOAL},i}/\pi_i(U))$) to find the best match (Kullback & Leibler, 1951); see **Figure 5c**. Once the voltages are obtained, measurements are taken again at the optimal voltages (vertically dashed lines in **Figure 5c**) that realize the respective gate (stars in **Figure 5a**).

The actual experimentally obtained distributions for the 0-clamped, 1-clamped, and also the unclamped case are shown in **Figure 5d, e** and **f**. The gate shows good performance, with a KLD of below 0.1 for the 1-clamped and unclamped distribution. For the 0-clamped distribution, the quality was bounded by the chosen ultra-low power voltage range of (-6.5 mV, 6.5mV) and could be improved by considering larger voltages, as the skyrmion is then always pushed into the (single) pinning site closest to a contact. Comparing with recent experimental s-MTJ implementations, our performance is very competitive (Zhang et al., 2025) and the KLD could be further improved by optimizing the gate positions. The unclamped configuration shows that the system can sample all valid logical states equally when inputs and outputs are not fixed, demonstrating its versatility. Prior work has shown that these kinds of systems also yield the correct output when both inputs are clamped, confirming their reliability for both directions of logic operations (Raab et al., 2022).

Our measurements show that skyrmion-based inverted logic gates offer a potential solution to combinatorial optimization problems and cryptographic tasks where invertible logic is essential (Aadit et al., 2021; Camsari et al., 2017; Smithson et al., 2019).

# Summary and Outlook

To summarize, we presented to our knowledge the first realization of an adaptable multi-value logic device prototype based on stochastic skyrmion dynamics. We build a simple coarse-grained model of linearly tunable effective pinning site energies to describe the stationary distribution of a confined skyrmion under the influence of spin-orbit torques. Direct access to multiple, tunable states enables the implementation of a scalable probabilistic multi-value computing system. In our system we demonstrated two key functionalities, firstly invertible logic OR gates by mapping the respective truth table entries to pinning sites. We were able to tune a single system to represent the 1-clamped, 0-clamped, and unclamped distribution by only varying the applied voltage. Secondly, we exploited similarities between the softmax function and the discrete Boltzmann law and demonstrated, based on MC simulations, that a neural network can be trained using a diffusion-based softmax approximation. As shown in previous Brownian skyrmion reservoir systems, the power consumption can be in the order of pW, allowing for ultra-low power operation mode (Beneke et al., 2024; Raab et al., 2022). With appropriate material design, the system can be scaled down into the sub-micron regime with the potential to tune the absolute timescales of the skyrmion dynamics, thus potentially increasing sampling speed (Miltat et al., 2018).

The ability to locally tune the energy of pinning sites significantly increases the flexibility of such a skyrmion-based MPC. Global influences are already used here, such as applying a DC current between two contacts. However, one could imagine adding more complex current patterns to vary the global scaling of the energies. Each new controllable current path would lead to an additional energy contribution $E_i(U) = E_i^0 + c_{A,i} \cdot U_A + c_{B,i} \cdot U_B + \dots$. Electrodes could also be connected from the top, at the center of a pinning site, to make the effect of voltage application more local (as the highest scaling is usually expected close to the contact, approximating the form $E_i(U) = E_i^0 + c_i \cdot U_i$ with maximum independent control of each pinning site). We want to point out that one does not have to rely on the naturally occurring pinning sites. Recently, it was also shown that focused ion beam as well as laser irradiation can influence skyrmion pinning locally, as the anisotropy change in the irradiated area produces new or enhances existing pinning sites in the material (Kern et al., 2022; He et al., 2023). This allows for full flexibility of the pinning sites to optimize these for a certain problem. Further, a constant magnetic field offset changes the skyrmion size and, therefore also, the effective pinning landscape (Gruber et al., 2022). And finally, one can even tune the pinning sites on-the-fly, as electrical fields have also been used to manipulate skyrmions locally using gates (Schott et al., 2017). An additionally applied alternating field generally increases the diffusion by up to two orders of magnitude (Gruber et al., 2023). Similar dynamics-enhancing effects can also be induced by random oscillating spin-orbit torques (Brems et al., 2021).

Our experimental MPC realization, presented here with one skyrmion $n_{sk}=1$ and $n_P=6$ pinning sites, has $K=n_p=6$ states. However, the number of states increases in a combinatorial fashion with the number of skyrmions and the number of available pinning sites. That provides the opportunity to scale the number of states to sizes relevant for applications.

A CMOS-compatible implementation of the skyrmion detection requires electrical read-out, e.g., by using magnetic tunnel junctions patterned on top of each pinning site and promising realizations are currently developed (Zhao et al., 2024).

Future research will explore the possibilities of local tunability as well as an upscaling of the system in order to tackle some of the most relevant problems, such as a max-SAT problem. For connecting multiple skyrmion-based invertible logic gates, the applied voltage of each gate needs to be controlled by the skyrmion position of the previous one (going backwards in the logic circuit). Also, the embedding of an experimental softmax function approximator would be a relevant next step towards applicability. So the scope for future developments based on this first proof-of-concept for MPC is very broad boding well for developing this further.

# Methods

## Clustering algorithm

For the development of a proper clustering algorithm, we took into account the following considerations: Since we are modeling the system as a Markovian process, clustering into pinning sites should fulfill properties of a reasonably good statistical clustering, such as the Perron-cluster cluster analysis *PCCA+* algorithm (Röblitz et al., 2013). This algorithm typically coarse-grains the transition matrix of microstates $\Lambda^{\kappa \times \kappa}$ such that coarse-grained macrostates (pinning sites in our case) have maximum self-transition rates, i.e. a high crispiness $\xi = \mathrm{trace}(t)/K$ of the coarse-grained transition matrix $t$. However, the methods of statistical physics operate on the transition matrix without knowledge of spatial information of the microstates, which can lead to disjoint clustering of microstates when projected back to the occurrence map.

On the other hand, a simple spatial clustering algorithm, such as *k-means* (Lloyd, 1982) on the positional data, with a coarse clustering $K$ (in our case, the number of pinning sites), produces joint clusters by definition. One might achieve reasonable values for the stationary distribution, however, it does not maximize self-transitions and thus might not converge to a desired solution.

A second challenge occurs during the data evaluation: Due to the limited image quality of Kerr microscopy data, the tracking mechanism fails from time to time (we approximate the failure rate in the order of ~1 % of the images). As this distorts the occurrence map as well as the dynamics evaluated, we have to filter out wrong labels as reliably as possible by our algorithm.

We aimed to implement a clustering algorithm which combines an iterative statistical coarse-graining with spatial information about the location of microstates and their neighbors, given by a neighborhood matrix $\underline{N}$ and a distance matrix $\underline{D}$, with $\underline{N}_{ij} \in \{0,1\}$ depending on whether two microstates are neighbors or not, and $\underline{D}_{ij} = |r_i - r_j|$ the absolute distance between two microstates, with $r_i$ the position of microstate $i$.

To coarse-grain each position of the skyrmion trajectory to a pinning site, we proceed as follows: We first discretize the trajectory into $\kappa_{init}$ microstates (we choose $\kappa_{init}=200$) with the *kmeans++* algorithm from the Python *scikit-learn* package (Arthur *et al.*, 2007), see **Figure 6** & also the small blue points in **Figure 2b** in the main text. To obtain pinning sites that are consistent throughout different measurements, we concatenate all trajectories as a set for collective clustering. We build a count matrix $\underline{C}_{init}$, an occurrence vector $O$, counting transitions between microstates. We further build a neighborhood matrix $\underline{N}_{init}$ by Voronoi tessellation/Delaunay triangulation and a distance matrix $\underline{D}_{init}$. We now filter out transitions and microstates based on the following conditions:

- Filter out transitions between microstates, which have a distance larger the $d_{max}$. These transitions are expected to occur predominantly due to mislabeling.
- Filter out microstates with has an overall occurrence $o < o_{min}$. We assume insufficient statistics for loosely occupied states.
- Filter out microstate transitions $C_{\zeta\eta} < c_{lim}$. We assume insufficient statistics for these transition.
- Filter out microstate transitions $C_{\zeta\eta}$, if $C_{\eta\zeta}=0$, with $C_{\zeta\eta}$ the counts from microstate $\zeta$ to $\eta$
- Delete microstates which has no occurrence at all after filtering events according to the filter rules to obtain new count matrix, neighborhood matrix and distance matrix $\underline{C}, \underline{N}, \underline{D}$ with number of microstates $\kappa \leq \kappa_{init}$.

We choose the following threshold values $d_{max}=0.2, c_{lim}=1, o_{lim}=3$ for the above algorithm. We start coarse-graining the system by assigning left-over microstates $\kappa$ to pinning sites $K$, following our motivation from above. As this is an iterative approach, we initialize the assignments ($\kappa \to K$) based on a *k-means++* algorithm on the raw skyrmion occurrence map. We construct a coarse-grained transition matrix $t_{init}^{K \times K}$. Now we start to iteratively improve the mapping with the idea to maximize self-transitions of the coarse-grained states. Concretely, we try to improve the crispiness $\xi$ of the clustering.

a. Pick randomly a microstate at the current pinning site boundaries. Calculate the crispiness $\xi_{now}$

b. (Hypothetically) re-assign the chosen microstate to the (all) neighbored pinning sites, given by the neighborhood matrix and the current assignment. Check the crispiness $\xi_{try}$.

c. If $\xi_{try} > \xi_{now}$, re-assign the chosen microstate, otherwise reject the change.

d. Repeat a-c $n$-times, choose $n$ large enough, such that crispiness $\xi$ converges. As we only re-assign microstates to neighboring pinning sites, we prevent disjoint clusters, while at the same time maximizing the self-transition rates under the condition of joint clusters.

Now separate the trajectories into sets of same stimulus (voltage) and reuse the obtained mappings $(x,y) \to \kappa_{init}$ and assignments $(\kappa \to K)$.

We want to note that for constructing count matrices or transition matrices, we always used a lag time $\tau = 1$, i.e. we evaluated transition between two consecutive Kerr microscopy images, with a frame rate of 16 Hz. **Figure 6** shows the initial (**a**) and final (**b**) assignment of microstates of the first skyrmion device prototype presented in the main text, as well as the crispiness increasing during the iterative clustering approach (**c**). The clustering algorithm runs for 1000 steps, while only steps are plotted in case $\xi$ increases.

## Numerical minimization of the energy model

We numerically minimize our problem of energy assignments to each pinning site, using the "Powell" method (Powell, 1964), implemented in the *scipy optimize* Python package module (Virtanen et al., 2020). Physically meaningful solutions deliver positive and negative slope values $c$, dependent on the relative position of the pinning sites to the contacts. We are therefore providing initial values for the minimization problem, as we recognized that the minimization process might lead to solutions that can not be interpreted in a physical way (e.g., all slopes are positive or negative, mainly due to the freedom of choosing an integration constant $G$, in our case $G(U)$).

For initial slope values $c_{\text{init}}$, we first perform current density path simulations for a (generic) material and normalize the current density $j$ to $j_{\text{NORM}}$ in the finite difference methods, using micmag2 simulation code (Litzius et al., 2024) and project the pinning site center points $x$ onto the mesh. We expect the slopes are scaling proportional to $(c_{\text{sim}})_i = \int_{x_0}^{r} j_{\text{NORM}} dr_s = (F_1)_i \cdot (c_{\text{init}})_i$. Here, $r_s$ is the streamline of the current density where the pinning site center $r$ sits, and $x_0$ is the tipping point on the streamline (In the 1D case of a homogeneous current density $j$, the formula simplifies to $(c_{\text{sim}})_i = (j \cdot d)_i$ with $d = x - x_0$ being the distance vector between pinning sites and center point). The current path formulas are numerically integrated, using appropriately small integration steps and interpolation of the discretized current density, which was performed using a finite difference scheme. **Figure 7a** shows the current density of the sample, illustrated by streamlines, and the tipping line lies between the two contacts, on which we assume a scaling of $c_{\text{init}} = 0$. Initial values of the stationary distribution $(E_{\text{init}})_i = (E_{\text{exp}} + F_2)_i = -\ln((\pi_{\text{exp}})_i) + G$ are obtained from experimental data as follows. First, minimize the auxiliary problem: a set of $(F_1, F_2)$ that minimizes

$E_{\text{init}} + c_{\text{init}} \cdot U_m = -\ln(\pi_{\text{exp}}(U)) + G(U)$, with $U_m$ being one of the measured voltages (in the case of the first device prototype $U_m \epsilon \{-1.5\,\text{mV}, -1\,\text{mV}, 0\,\text{mV}, 1.5\,\text{mV}, 2\,\text{mV}\}$, for the second device prototype $U_m \epsilon \{-2\,\text{mV}, -1\,\text{mV}, 0\,\text{mV}, 1\,\text{mV}, 2\,\text{mV}\}$). We set the constraint that the offset value of the 0 mV-measurement are set to 0, $F_2^{0\,\text{mV}}=0$. The loss for the auxiliary problem was defined as

$L_{\text{aux}}=\sum_{k=1}^{K}\left|E_{\text{init},k}+c_{\text{init},k}\cdot U+\ln\left(\pi_{\exp,k}(U)\right)\right|$. The loss function for the minimization of the actual

problem is defined as $L=\sum_{m}\sum_{k=1}^{K}\left(\pi_{\exp_k}(U_m)-\pi_{\text{fit}_k}(U_m)\right)^2$ with $\pi_{\text{fit}}(U_m)=e^{-\left(E_k^0+c_k\cdot U_m\right)}/\sum e^{-\left(E_k^0+c_k\cdot U_m\right)}$.

**Figure 7b** shows the initial values $(E_{\text{init}},c_{\text{init}})$ used for minimization as lines.

## Experimental details

Skyrmion motion is detected using a magneto-optical Kerr effect (MOKE) microscope manufactured by evico magnetics GmbH.

The microscope is equipped with a CMOS camera providing a temporal resolution of 16 frames per second, with an exposure time of 62.5 ms. To enhance the signal-to-noise ratio, binning is applied, reducing the effective resolution as a trade-off for better contrast and less storage needed for long-time measurements. To improve the image contrast, differential imaging is employed by subtracting the saturated state image from the skyrmion state. For the tracking procedure, the *trackpy* package was used (Allan et al., 2024).

The magnetic field environment is controlled by a set of custom-designed coils: one coil generates the out-of-plane (OOP) magnetic field, while a pair of coils provides the in-plane (IP) field. A Peltier element is coupled to the OOP coil, enabling temperature control within the range of 285–360 K, with a thermal stability of ±0.3 K. During the experiments, the system was maintained at a constant temperature around room temperature, as monitored by a Pt100 resistive temperature sensor. To further enhance thermal stability, the entire microscope system is housed within a laminar flow enclosure equipped with precise temperature regulation. Additionally, a piezoelectric stage is coupled to the OOP coil to compensate for drift effects. This integrated setup ensures minimal thermal and mechanical drift, enabling stable, long-duration measurements at high magnification.

A CoFeB magnetic layer is used, embedded in a stack of several nonmagnetic layers: [Ta(5)/Co$_{20}$Fe$_{60}$B$_{20}$(0.95)/Ta(0.09)/MgO(2)/Ta(5)]. The thickness of each layer is given in nanometers in brackets. **Table 2** provides detailed information on measurement details.

Skyrmion nucleation is initiated by applying an in-plane (IP) magnetic field pulse , in the presence of a stabilizing out-of-plane (OOP) bias field. Following nucleation, oscillations in the OOP field are introduced to increase skyrmion diffusion while maintaining the bias field. As the input signal, a constant voltage is applied at two opposing corners (top and bottom right) of the samples using a Keithley 2400 source meter. To enhance statistical reliability, measurements are performed for at least 45 minutes at each voltage setting.

# **Acknowledgment**

T.B.W, J.H.M. and M.K acknowledge funding from the European Union's Horizon 2020 Research and Innovation Programme under Grant Agreement No 856538 (ERC-2019-SyG) and the Horizon Europe project no. 101070290 (NIMFEIA). This project has received funding from the European Union's Horizon Europe Programme Horizon.1.2 under the Marie Skłodowska-Curie Actions (MSCA), Grant agreement No. 101119608 (TOPOCOM). J.H.M. acknowledges funding VIDI talent programme of the Dutch Research Council (NWO), project 223.157 (CHASEMAG) and KIC project no. 22016. M.C. and D.R. acknowledge support by the Contract n. 2025-40-I.0 (SPINAM) funded by the Italian Space Agency within the call "Studi di concetti innovativi di sistemi spaziali". The work in Mainz (G.B., F.K. and M.K.) was supported by the Deutsche Forschungsgemeinschaft (DFG, German Research Foundation) projects 403502522 (SPP 2137 Skyrmionics), 49741853, and 268565370 (SFB TRR173 projects A01, B02 and A12) as well as TopDyn and the Zeiss foundation through the Center for Emergent Algorithmic Intelligence. This research was supported by the National Research Council of Science & Technology (NST) grant by the Korean government MSIT (Grant No. GTL24041-000). The work is a highly interactive collaboration supported by the Horizon 2020 Framework Program of the European Commission under FET-Open grant agreement no. 863155 (s-Nebula).

# Bibliography

Aadit, N. A., Grimaldi, A., Carpentieri, M., Theogarajan, L., Finocchio, G., & Camsari, K. Y. (2021). Computing with Invertible Logic: Combinatorial Optimization with Probabilistic Bits. *2021 IEEE International Electron Devices Meeting (IEDM)*, 40.3.1-40.3.4. https://doi.org/10.1109/IEDM19574.2021.9720514

Allan, D. B., Caswell, T., Keim, N. C., van der Wel, C. M., & Verweij, R. W. (2024). soft-matter/trackpy: V0.6.4 (Version v0.6.4) [Computer software]. Zenodo. https://doi.org/10.5281/ZENODO.1213240

Andreev, M., Seo, S., Jung, K., & Park, J. (2022). Looking Beyond 0 and 1: Principles and Technology of Multi-Valued Logic Devices. *Advanced Materials*, *34*(51), 2108830. https://doi.org/10.1002/adma.202108830

Arthur, D. & Vassilvitskii, S. (2007). k-means++: the advantages of careful seeding, *SODA '07: Proceedings of the eighteenth annual ACM-SIAM symposium on Discrete algorithms, 1027 – 1035.* http://dx.doi.org/10.1145/1283383.1283494

Banerjee, K., C, V. P., Gupta, R. R., Vyas, K., H, A., & Mishra, B. (2020). *Exploring Alternatives to Softmax Function* (Version 1). arXiv. https://doi.org/10.48550/ARXIV.2011.11538

Beaumont, M. A. (2019). Approximate Bayesian Computation. *Annual Review of Statistics and Its Application*, *6*(1), 379–403. https://doi.org/10.1146/annurev-statistics-030718-105212

Bednarz, B., Syskaki, M.-A., Pachat, R., Prädel, L., Wortmann, M., Kuschel, T., Ono, S., Kläui, M., Herrera Diez, L., Jakob, G. (2024). Stabilizing perpendicular magnetic anisotropy with strong exchange bias in PtMn/Co by magneto-ionics *Appl. Phys. Lett.* **124**, 232403. https://doi.org/10.1063/5.0213731

Beneke, G., Winkler, T. B., Raab, K., Brems, M. A., Kammerbauer, F., Gerhards, P., Knobloch, K., Mentink, J., & Kläui, M. (2024). Gesture recognition with Brownian reservoir computing using geometrically confined skyrmion dynamics. *Nature Communications* 15, 8103. https://doi.org/10.1038/s41467-024-52345-y

Borders, W. A., Pervaiz, A. Z., Fukami, S., Camsari, K. Y., Ohno, H., & Datta, S. (2019). Integer factorization using stochastic magnetic tunnel junctions. *Nature*, *573*(7774), 390–393. https://doi.org/10.1038/s41586-019-1557-9

Brems, M. A., Kläui, M., & Virnau, P. (2021). Circuits and excitations to enable Brownian token-based computing with skyrmions. *Applied Physics Letters*, *119*(13), Article 13. https://doi.org/10.1063/5.0063584

Camsari, K. Y., Faria, R., Sutton, B. M., & Datta, S. (2017). Stochastic p -Bits for Invertible Logic. *Physical Review X*, *7*(3), 031014. https://doi.org/10.1103/PhysRevX.7.031014

Capra, M., Bussolino, B., Marchisio, A., Masera, G., Martina, M., & Shafique, M. (2020). Hardware and Software Optimizations for Accelerating Deep Neural Networks: Survey of Current Trends, Challenges, and the Road Ahead. *IEEE Access*, *8*, 225134–225180. https://doi.org/10.1109/ACCESS.2020.3039858

Celikyilmaz, A., Türksen, I.B. (2009). Modeling Uncertainty with Fuzzy Logic, *Springer Berlin*, Heidelberg, ISBN 978-3-540-89923-5, https://doi.org/10.1007/978-3-540-89924-2

Debashis, P., Li, H., Nikonov, D., & Young, I. (2021). Gaussian Random Number Generator with Reconfigurable Mean and Variance using Stochastic Magnetic Tunnel Junctions. *ArXiv* 2112.04577. https://doi.org/10.48550/ARXIV.2112.04577

De Vries, Alex (2023). The growing energy footprint of artificial intelligence. Joule, 7 (10), 2191-2194. https://doi.org/10.1016/j.joule.2023.09.004

Duffee, C., Athas, J., Shao, Y., Melendez, N. D., Raimondo, E., Katine, J. A., Camsari, K. Y., Finocchio, G., & Khalili Amiri, P. (2025). An integrated-circuit-based probabilistic computer that uses voltage-controlled magnetic tunnel junctions as its entropy source. *Nature Electronics*. https://doi.org/10.1038/s41928-025-01439-6

Finocchio, G., Incorvia, J. A. C., Friedman, J. S., Yang, Q., Giordano, A., Grollier, J., Yang, H., Ciubotaru, F., Chumak, A. V., Naeemi, A. J., Cotofana, S. D., Tomasello, R., Panagopoulos, C., Carpentieri, M., Lin, P., Pan, G., Yang, J. J., Todri-Sanial, A., Boschetto, G., … Bandyopadhyay, S. (2024). Roadmap for unconventional computing with nanotechnology. *Nano Futures*, *8*(1), 012001. https://doi.org/10.1088/2399-1984/ad299a

Fisher, R.A. (1936). The use of multple measurements in taxonomic problems, Annals of Eugenics, 7 (2), 179-188. https://doi.org/10.1111/j.1469-1809.1936.tb02137.x

Gawlikowski, J., Tassi, C. R. N., Ali, M., Lee, J., Humt, M., Feng, J., Kruspe, A., Triebel, R., Jung, P., Roscher, R., Shahzad, M., Yang, W., Bamler, R., & Zhu, X. X. (2023a). A survey of uncertainty in deep neural networks. *Artificial Intelligence Review*, *56*(S1), 1513–1589. https://doi.org/10.1007/s10462-023-10562-9

Grimaldi, A., Sánchez-Tejerina, L., Anjum Aadit, N., Chiappini, S., Carpentieri, M., Camsari, K., & Finocchio, G. (2022). Spintronics-compatible Approach to Solving Maximum-Satisfiability Problems with Probabilistic Computing, Invertible Logic, and Parallel Tempering. *Physical Review Applied*, *17*(2). https://doi.org/10.1103/physrevapplied.17.024052

Gruber, R., Brems, M. A., Rothörl, J., Sparmann, T., Schmitt, M., Kononenko, I., Kammerbauer, F., Syskaki, M., Farago, O., Virnau, P., & Kläui, M. (2023). 300-Times-Increased Diffusive Skyrmion Dynamics and Effective Pinning Reduction by Periodic Field Excitation. *Advanced Materials*, *35*(17), Article 17. https://doi.org/10.1002/adma.202208922

Gruber, R., Zázvorka, J., Brems, M. A., Rodrigues, D. R., Dohi, T., Kerber, N., Seng, B., Vafaee, M., Everschor-Sitte, K., Virnau, P., & Kläui, M. (2022). Skyrmion pinning energetics in thin film systems. *Nature Communications*, *13*(1), Article 1. https://doi.org/10.1038/s41467-022-30743-4

Han, X., Zhu, X., Pedrycz, W., Mostafa, A. M., & Li, Z. (2024). A design of fuzzy rule-based classifier optimized through softmax function and information entropy. *Applied Soft Computing*, *156*, 111498. https://doi.org/10.1016/j.asoc.2024.111498

He, B., Tomasello, R., Luo, Xuming, Zhang, R., Zhuyang, N., Carpentieri, M., Han, X., Finocchio, G., Yu, G. (2023). All-Electrical 9-Bit Skyrmion-Based Racetrack Memory Designed with Laser Irradiation. *Nano Letters* 23(20), 9482-9490. https://doi.org/10.1021/acs.nanolett.3c02978

He, Y.-L., Zhang, X.-L., Ao, W., & Huang, J. Z. (2018). Determining the optimal temperature parameter for Softmax function in reinforcement learning. *Applied Soft Computing*, *70*, 80–85. https://doi.org/10.1016/j.asoc.2018.05.012

Husic, B.E. and Pande, V.S. 2018. Markov State Models: From an Art to a Science. *J. Am. Chem. Soc.* **140**, 2386-2396. http://dx.doi.org/10.1021/jacs.7b12191

Jaber, R. A., El-Hajj, A. M., Kassem, A., Nimri, L. A., & Haidar, A. M. (2020). CNFET-based designs of Ternary Half-Adder using a novel "decoder-less" ternary multiplexer based on unary operators. *Microelectronics Journal*, *96*, 104698. https://doi.org/10.1016/j.mejo.2019.104698

Jia, X., Yang, J., Dai, P., Liu, R., Chen, Y., & Zhao, W. (2020). SPINBIS: Spintronics based Bayesian Inference System with Stochastic Computing. *IEEE Transactions on Computer-Aided Design of Integrated Circuits and Systems*, *39*(4), 789–802. https://doi.org/10.1109/TCAD.2019.2897631

Kern, L.-M., Pfau, B., Deinhart, V., Schneider, M., Klose, C., Gerlinger, K., Wittrock, S., Engel, D., Will, I., Günther, C. M., Liefferink, R., Mentink, J. H., Wintz, S., Weigand, M., Huang, M.-J., Battistelli, R., Metternich, D., Büttner, F., Höflich, K., & Eisebitt, S. (2022). Deterministic Generation and Guided Motion of Magnetic Skyrmions by Focused He + -Ion Irradiation. *Nano Letters*, *22*(10), Article 10. https://doi.org/10.1021/acs.nanolett.2c00670

Kullback, S. & Leibler, R.A. (1951). On Information an Sufficiency. Ann. Math. Statist. 22(1), 79-86. https://doi.org/10.1214/aoms/1177729694

Lukasiewicz, T. (1999). Probabilistic and truth-functional many-valued logic programming. *Proceedings 1999 29th IEEE International Symposium on Multiple-Valued Logic (Cat. No.99CB36329)*, 236–241. https://doi.org/10.1109/ISMVL.1999.779722

Maharjan, S., Alsadoon, A., Prasad, P. W. C., Al-Dalain, T., & Alsadoon, O. H. (2020). A novel enhanced softmax loss function for brain tumour detection using deep learning. *Journal of Neuroscience Methods*, *330*, 108520. https://doi.org/10.1016/j.jneumeth.2019.108520

Miltat, J., Rohart, S., & Thiaville, A. (2018). Brownian motion of magnetic domain walls and skyrmions, and their diffusion constants. *Physical Review B*, *97*(21), 214426. https://doi.org/10.1103/PhysRevB.97.214426

Nikhar, S., Kannan, S., Aadit, N. A., Chowdhury, S., & Camsari, K. Y. (2024). All-to-all reconfigurability with sparse and higher-order Ising machines. *Nature Communications*, *15*(1), 8977. https://doi.org/10.1038/s41467-024-53270-w

Onizawa, N., Nishino, K., Smithson, S. C., Meyer, B. H., Gross, W. J., Yamagata, H., Fujita, H., & Hanyu, T. (2021). A Design Framework for Invertible Logic. *IEEE Transactions on Computer-Aided Design of Integrated Circuits and Systems*, *40*(4), 655–665. https://doi.org/10.1109/TCAD.2020.3003906

Parr, T., Rees, G., & Friston, K. J. (2018). Computational Neuropsychology and Bayesian Inference. *Frontiers in Human Neuroscience*, *12*, 61. https://doi.org/10.3389/fnhum.2018.00061

Pinto, J. P., Kelur, S., & Shetty, J. (2018). Iris Flower Species Identification Using Machine Learning Approach. *2018 4th International Conference for Convergence in Technology (I2CT)*, 1–4. https://doi.org/10.1109/I2CT42659.2018.9057891

Powell, M. J. D. (1964). An efficient method for finding the minimum of a function of several variables without calculating derivatives. *Computer Journal.* **7** (2): 155–162. doi:10.1093/comjnl/7.2.155. hdl:10338.dmlcz/103029

Raab, K., Brems, M. A., Beneke, G., Dohi, T., Rothörl, J., Kammerbauer, F., Mentink, J. H., & Kläui, M. (2022). Brownian reservoir computing realized using geometrically confined skyrmion dynamics. *Nature Communications*, *13*(1), Article 1. https://doi.org/10.1038/s41467-022-34309-2

Raimondo, E., Garzòn, E., Shao, Y., Grimaldi, A., Chiappini, S., Tomasello, R., Davila-Melendez, N., Katine, J. A., Carpentieri, M., Chiappini, M., Lanuzza, M., Pedram Khalili, A., Finocchio, G. (2025). High-performance and reliable probabilistic Ising machine based on simulated quantum annealing. *Arxiv 2503.13015*. https://doi.org/10.48550/arXiv.2503.13015

Reichhardt, C., Reichhardt, C.J.O. (2016). Noise Fluctuations and Drive Dependence of the Skyrmion Hall Effect in Disordered Systems. *New. J. Phys.* **18**, 095005. https://doi.org/10.1088/1367-2630/18/9/095005

Rohmer, J. (2020). Uncertainties in conditional probability tables of discrete Bayesian Belief Networks: A comprehensive review. *Engineering Applications of Artificial Intelligence*, *88*, 103384. https://doi.org/10.1016/j.engappai.2019.103384

Röblitz, S. & Weber, M. (2013). Fuzzy spectral clustering by PCCA+: application to Markov state models and data classification. *Adv. Data Anal. Classif. 7, 147-179.* https://doi.org/10.1007/s11634-013-0134-6

Rue, H., Riebler, A., Sørbye, S. H., Illian, J. B., Simpson, D. P., & Lindgren, F. K. (2017). Bayesian Computing with INLA: A Review. *Annual Review of Statistics and Its Application*, *4*(1), 395–421. https://doi.org/10.1146/annurev-statistics-060116-054045

Schäfer, R. (2007). Investigation of Domains and Dynamics of Domain Walls by the Magneto-optical Kerr-effect. *Handbook of Magnetism and Advanced Magnetic Materials*. John Wiley & Sons. https://doi.org/10.1002/9780470022184.hmm310

Schmidt, F., Rave, W., & Hubert, A. (1985). Enhancement of magneto-optical domain observation by digital image processing. *IEEE Transactions on Magnetics*, *21*(5), Article 5. https://doi.org/10.1109/TMAG.1985.1064048

Schott, M., Bernard-Mantel, A., Ranno, L., Pizzini, S., Vogel, J., Béa, H., Baraduc, C., Auffret, S., Gaudin, G. & Givord, D. (2017), The Skyrmion Switch: Tuning Magnetic Skyrmion Bubbles on and off with an Electric Field. *ACS Nano Letters* **17**, 3006-3012. https://pubs.acs.org/doi/abs/10.1021/acs.nanolett.7b00328

Singh, N.S., Delacour, C., Niazi, S., Selcuk, K., Golenchenko, D., Kaneko, H., Kanai, S., Ohno, H., Fukami, S., Camsari,K.Y., *(2024)*. Beyond Ising: Mixed Continuous Optimization with Gaussian Probabilistic Bits Using Stochastic MTJs, *IEEE International Electron Devices Meeting (IEDM)*, San Francisco, CA, USA, 2024, pp. 1-4, doi: https://10.1109/IEDM50854.2024.10873478.

Smithson, S. C., Onizawa, N., Meyer, B. H., Gross, W. J., & Hanyu, T. (2019). Efficient CMOS Invertible Logic Using Stochastic Computing. *IEEE Transactions on Circuits and Systems I: Regular Papers*, *66*(6), 2263–2274. https://doi.org/10.1109/TCSI.2018.2889732

Soh, K., Kim, J. E., Chun, S. Y., & Yoon, J. H. (2025). Calibration of P-bit for aligned stochastic outputs in probabilistic computing. *Materials Science and Engineering: B*, *317*, 118146. https://doi.org/10.1016/j.mseb.2025.118146

Swain, M. (2012). An Approach for IRIS Plant Classification Using Neural Network. *International Journal on Soft Computing*, *3*(1), 79–89. https://doi.org/10.5121/ijsc.2012.3107

Thiele, A. A. (1973). Steady-State Motion of Magnetic Domains. *Phys. Rev. Lett.* **30**, 230. https://doi.org/10.1103/PhysRevLett.30.230

Lin, S.-Z., Reichhardt, C., Batista, C. D. & Saxena, A. (2013). Particle model for skyrmions in metallic chiral magnets: dynamics, pinning, and creep. *Phys. Rev. B* **87**, 214419. https://doi.org/10.1103/PhysRevB.87.214419

Litzius K., Winkler, T.B. *et al. (2024),* micmag2, https://github.com/micmag2

Lloyd, S. (1982). Least squares quantization in PCM. *IEEE Transactions on Information Theory*. **28** (2): 129–137. doi:10.1109/TIT.1982.1056489

Vaswani, A., Shazeer, N., Parmar, N., Uszkoreit, J., Jones, L., Gomez, A. N., Kaiser, L., & Polosukhin, I. (2017). Attention Is All You Need (Version 7). *Arxiv, 1706.03762*. https://doi.org/10.48550/ARXIV.1706.03762

Virtanen, P., Gommers, R., Oliphant, T.E. *et al.* (2020). SciPy 1.0: fundamental algorithms for scientific computing in Python. *Nat Methods* **17**, 261–272 . https://doi.org/10.1038/s41592-019-0686-2

Whitehead, W., Nelson, Z., Camsari, K.Y. *et al.* CMOS-compatible Ising and Potts annealing using single-photon avalanche diodes. *Nat Electron* **6**, 1009–1019 (2023). https://doi.org/10.1038/s41928-023-01065-0

Wu, F. Y. (1982). The Potts model. *Reviews of Modern Physics*, *54*(1), 235–268. https://doi.org/10.1103/RevModPhys.54.235

Xia, Q., & Yang, J. J. (2019). Memristive crossbar arrays for brain-inspired computing. *Nature Materials*, *18*(4), 309–323. https://doi.org/10.1038/s41563-019-0291-x

Yang, S., Grimaldi, A., Bao., Youwei, Raomindo, E., Si, J., Finocchio, G., Yang, H. (2025). 250 Magnetic Tunnel Junctions-Based Probabilistic Ising Machine. *Arxiv*, 2506.14590. https://arxiv.org/pdf/2506.14590

Yoo, H., Chang-Hyun, K. "Multi-valued logic system: New opportunities from emerging materials and devices." *Journal of Materials Chemistry C* 9.12 (2021): 4092-4104. https://doi.org/10.1039/D1TC00148E

Zadeh, L. A. (1988). Fuzzy logic. *Computer*, *21*(4), 83–93. https://doi.org/10.1109/2.53

Zadeh, L. A. (1999). From computing with numbers to computing with words. From manipulation of measurements to manipulation of perceptions. *IEEE Transactions on Circuits and Systems I: Fundamental Theory and Applications*, *46*(1), 105–119. https://doi.org/10.1109/81.739259

Zadeh, L. A. (2023). Fuzzy logic. In *Granular, Fuzzy, and Soft Computing* (pp. 19–49). Springer US. https://doi.org/10.1007/978-1-0716-2628-3_234

Zhang, B., Liu, Y., Gao, T., Yin, J., Guan, Z., Zhang, D., Zeng, L. (2025). Automatic Extraction and Compensation of P-Bit Device Variations in Large Array Utilizing Boltzmann Machine Training. *Micromachines* **16** (2), 133. https://doi.org/10.3390/mi16020133

Zhao, M., Chen, A., Huang, P.-Y., Liu, C., Shen, L., Liu, J., Zhao, L., Fang, B., Yue, W.-C., Zheng, D., Wang, L., Bai, H., Shen, K., Zhou, Y., Wang, S., Liu, E., He, S., Wang, Y.-L., Zhang, X., & Jiang, W. (2024). Electrical detection of mobile skyrmions with 100% tunneling magnetoresistance in a racetrack-like device. *Npj Quantum Materials*, *9*(1), Article 1. https://doi.org/10.1038/s41535-024-00655-1

Zhu, D., Lu, S., Wang, M., Lin, J., & Wang, Z. (2020). Efficient Precision-Adjustable Architecture for Softmax Function in Deep Learning. *IEEE Transactions on Circuits and Systems II: Express Briefs*, *67*(12), 3382–3386. https://doi.org/10.1109/TCSII.2020.3002564

Zink, B. R., Lv, Y., Wang, J.-P. (2022). Review on stochastic magnetic tunnel junctions for stochastic computing. *IEEE journal on Exploratory Solid State Computational Devices and Circuits* **8** (2). https://doi.org/10.1109/JXCDC.2022.3227062

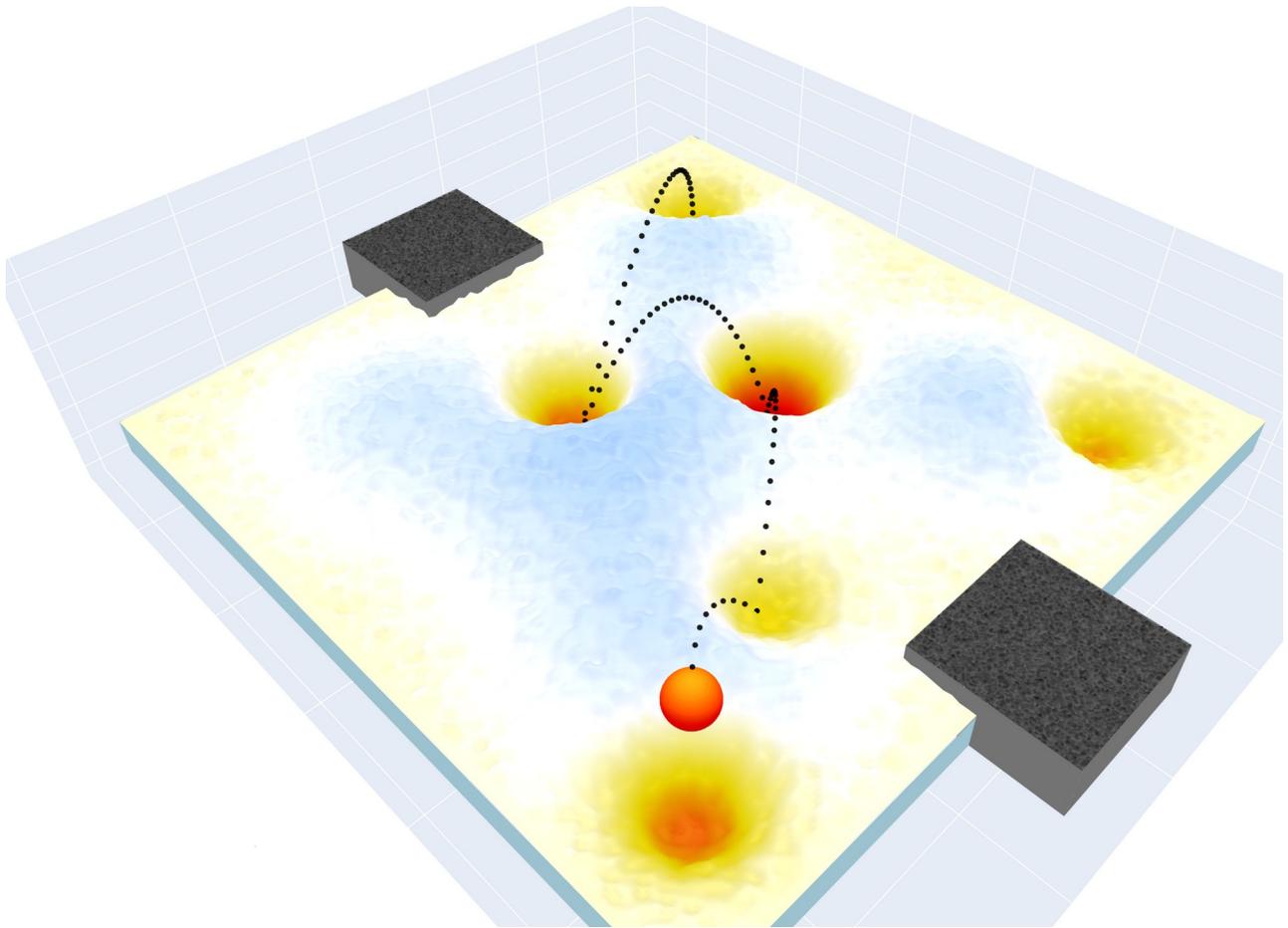

*Figure 1: Sketch of a device-specific effective energy landscape in real space, as felt by a magnetic skyrmion (here visualized as a red particle). Multiple pinning sites are indicated as local minima (white-orange-red areas). The skyrmion itself will be found in one of the pinning sites and eventually hop from one to another (black dotted lines). A voltage can be applied between the two electrodes (grey) to tune the energy landscape. The pinning sites are a natural source for a coarse-graining of the system into a few parameters.*

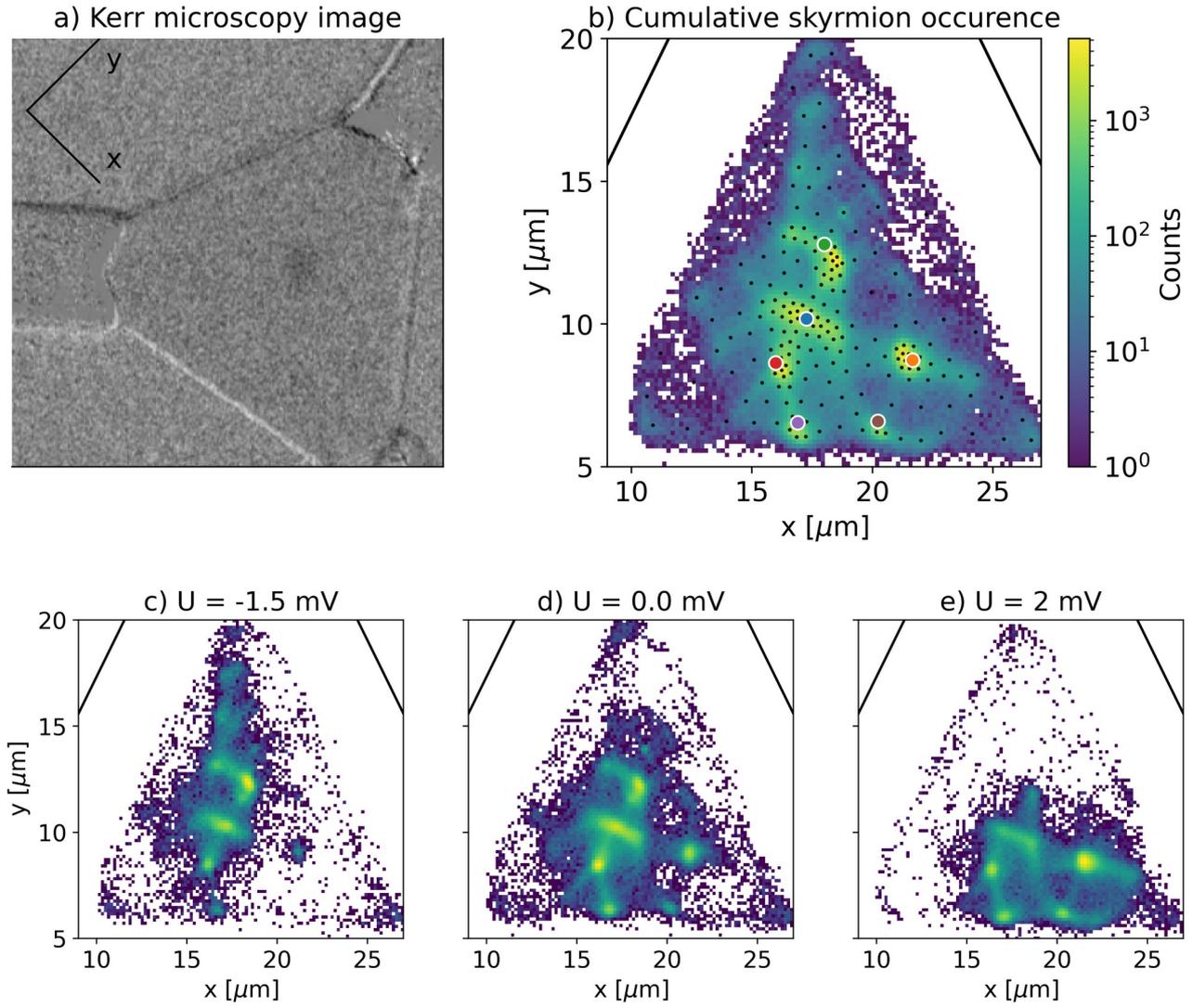

*Figure 2: a) Snapshot of the triangular device prototype, hosting a skyrmion (darker gray spot near the center). The gray scale contrast within the triangle corresponds to the out-of-plane magnetization component (bright – down, dark - up). b) Cumulative skyrmion occurrence from long-term measurements. We clearly see the $K=6$ pinning sites, indicated with colored discs. The color-coding of the pinning sites is also later used for **Figure 3**. Small dark blue points indicate the space discretization into microstates as an intermediate step (see Methods section). c) – e) Experimentally observed skyrmion occurrence map for different applied voltages $U$. The skyrmion gets pushed to one of the contacts, depending on the sign and strength of the applied voltage. Yellow refers to high occupancy, violet refers to low occupancy. **Figure 7a** shows the current simulated density path for the triangular sample.*

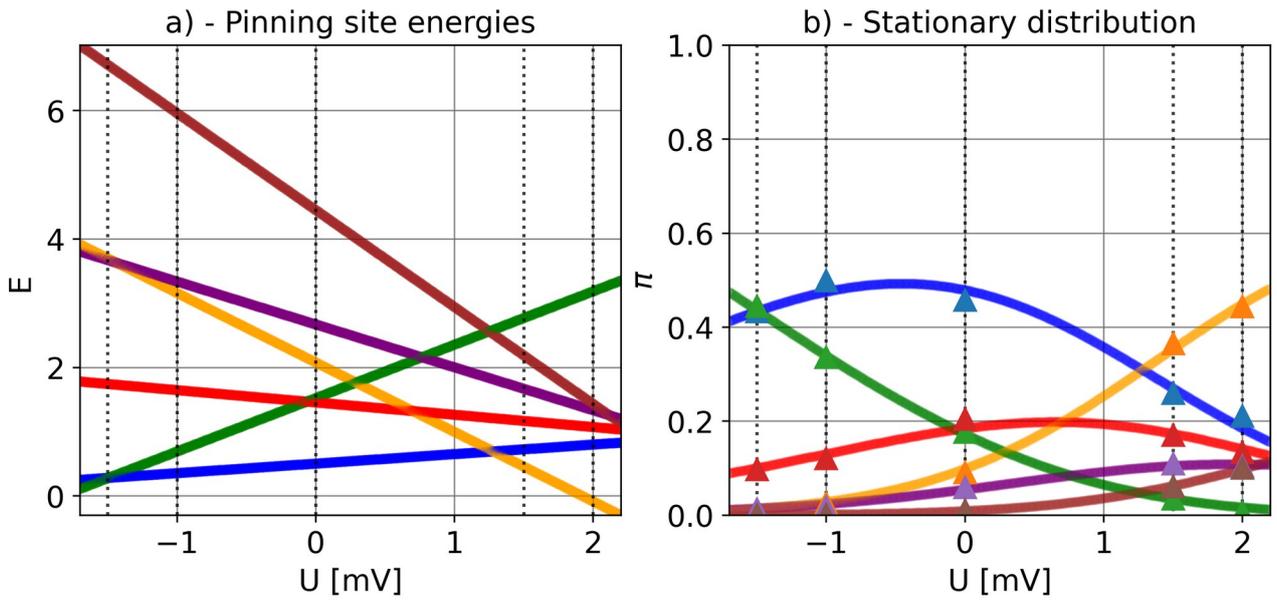

*Figure 3: a) The modeled energies and their linear scaling with applied voltage. The line colors correspond to the pinning site markers shown in **Figure 2b**. b) The modeled stationary distribution π and its voltage dependence, compared to experimentally observed data $\pi_{\text{exp}}$ (triangles), showing good agreement.*

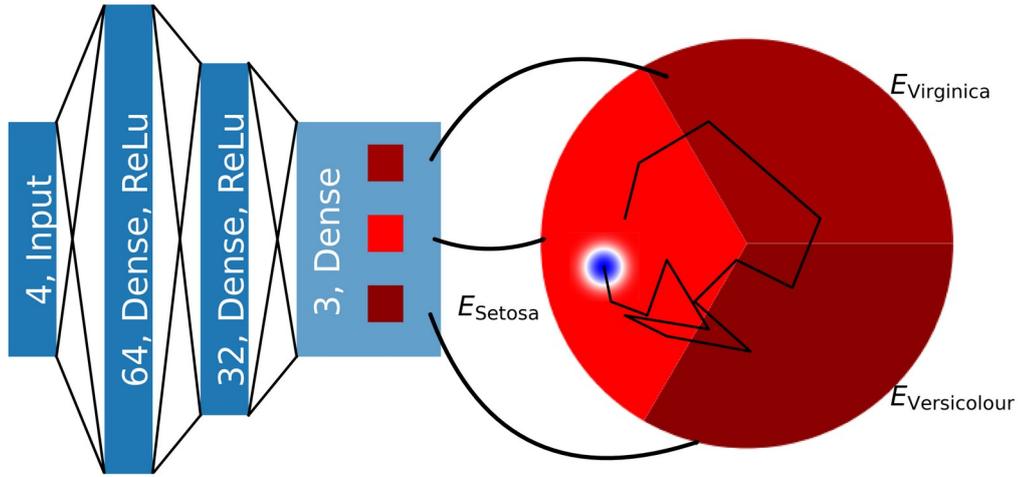
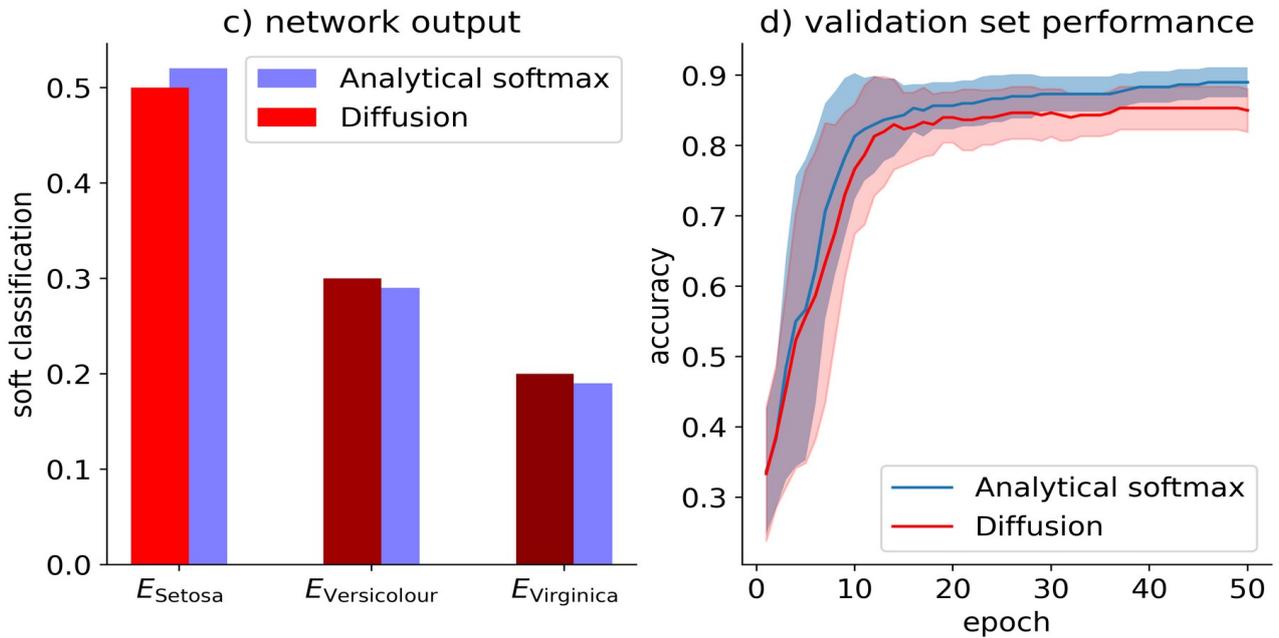

*Figure 4*: Diffusion-based softmax function for the classification of the Iris dataset, with three classes. a) network visualization. The negative pre-activations of the output layer are interpreted as energies in the plot. b) The diffusive skyrmion explores the energy landscape, split into three regions, with respective energies obtained from network inference. Results are obtained from Monte-Carlo simulations. c) The analytical softmax and the finite-time diffusive estimation are compared (for an arbitrary network state). We compare the performance of the values calculated analytically with the values obtained by simulation. d) Network performance on the validation set during training, comparison between analytical and diffusive softmax. Error bars indicate the mean deviation after 10 runs.

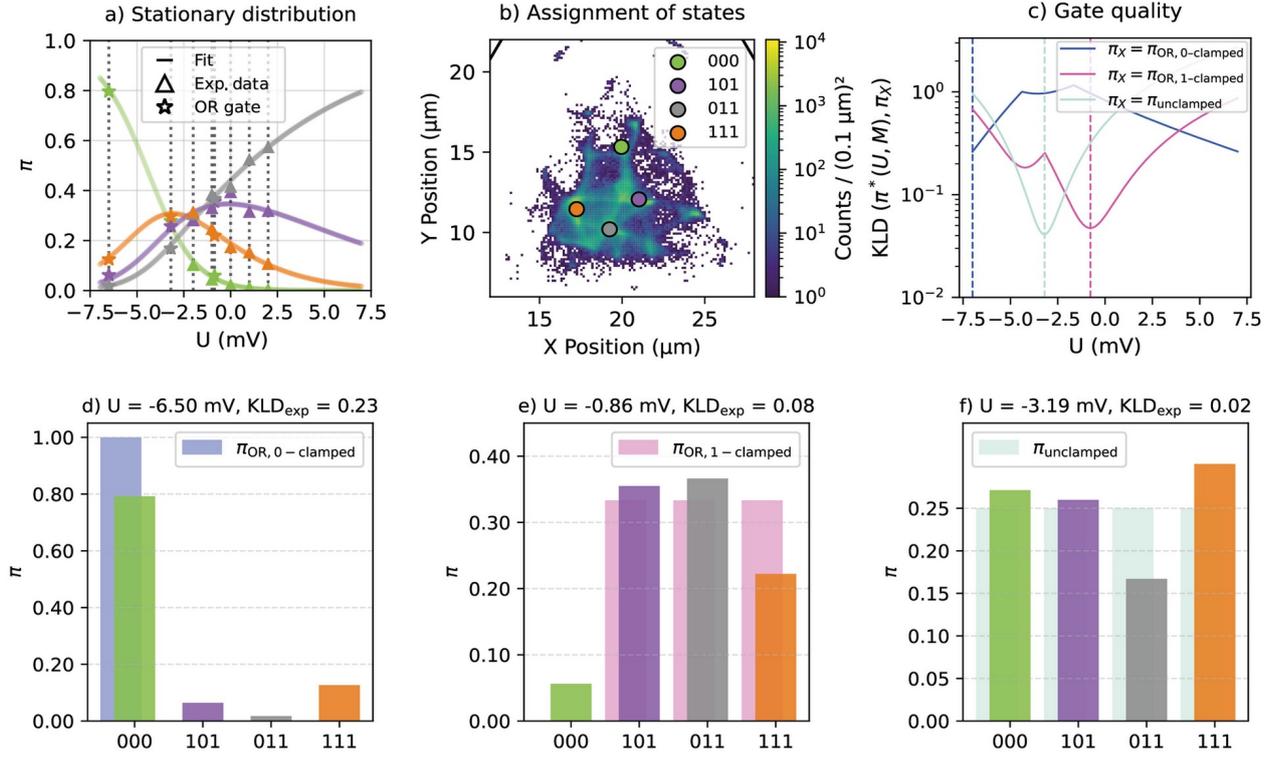

*Figure 5*: Invertible OR state with a second MPC device prototype. a) stationary distribution. b) Assignment of logical OR truth table entries to pinning sites. c) Kullback-Leibler divergence for the comparison between target distribution and experimentally realized distribution by scanning the voltage. d)-f) The best match of an experimentally measured stationary distribution (colors of pinning sites) with the "goal" distribution (uniform color, according to c)), according to KLD. The three numbers encode the logical variables in the order Input A/Input B/Output. We find a competitive performance in terms of accuracy. The x-axis tick labels encode the valid logical states of the OR gate; the numbers refer to (Input 1, Input 2, Output).

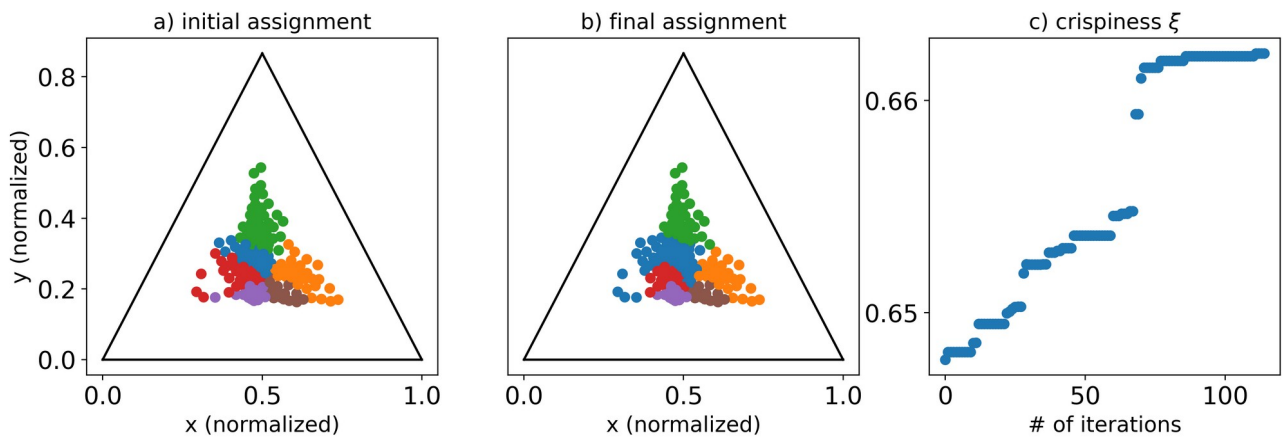

*Figure 6*: *The occurrence map initial (a) and final (b) assignment, and c) the increase of crispiness during the clustering algorithm. Main text figure 2b also shows the 200 microstates as small dark blue points. Comparing the shape of the final assignment with the occurrence map shows good clustering (joint areas of high probability).*

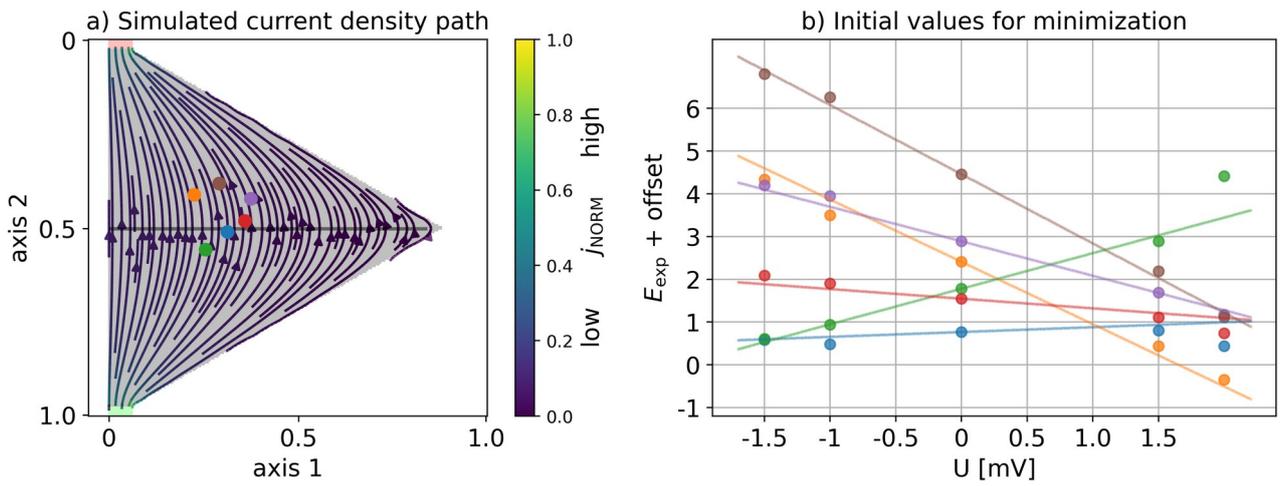

***Figure 7**: a) Normalized in-plane current density and direction in a (generic) triangular sample with contacts at two edges (the current density is color-coded). The current path is used to estimate initial values for the minimization process of the model. Without proper initial values, minimization might lead to equally valid, but not physically interpretable, solutions of the model parameters. Approximate pinning site centers are plotted, rotated in such a way that their position with respect to the electrodes is kept.*

*b) Auxiliary minimization problem to adjust scaling parameters obtained from normalized current density path to energies (negative log-probabilities) obtained from experiment ($-\log(\pi_{\exp})$). Data points show estimated energies from experimental data (plus fitted offset), while the lines are showing the best linear model fitting to experimental data, by fixing scaling factors obtained from current density simulations, only allowing single offsets for each voltage data has been gathered.*

| Input A | Input B | Output C |
|---------|---------|----------|
| 0 | 0 | 0 |
| 0 | 1 | 1 |
| 1 | 0 | 1 |
| 1 | 1 | 1 |

*Table 1*: *OR-gate truth table*

|  | 1st device<br>**Modeling** | 2nd device<br>**Invertible Logic** |
|---|---|---|
| Camera | Hamamatsu Digital CCD Camera (C8484-03G02) | DMK 37BUX250 |
| Binning | 2x2 | 4x4 |
| Spatial resolution | 672x512 | 612x512 |
| Temperature | 320 K | 313.3 K |
| IP pulse for skyrmion nucleation | 20 mT for 0.5 s | 30 mT, for1 s |
| Stabilizing out-of-plane (OOP) bias field | 50 µT | 100 µT |
| oscillatory field amplitude | 50 µT | 10 µT |
| oscillatory field frequency | 100 Hz | 10 Hz |

*Table 2*: *Details on experimental measurements*